\documentstyle[aps,amssymb]{revtex}

\begin{document}
\author{Nikolai P. Tretiakov\thanks{%
e-mail: nicolai@fis.ufg.br} and J.N.Teixeira Rabelo\thanks{%
e-mail: jrabelo@fis.ufg.br}}
\address{Instituto de F\'{i}sica, Universidade Federal de Goias, P.O. Box 131,\\
74001-970, Goiania, Goias, Brazil}
\title{Effective Liouville Equation for Classical Driven Systems\thanks{%
work supported by CNPq, Brazil}}
\date{10 June 1998 }
\maketitle

\begin{abstract}
A large class of classical dynamical systems with an external rapidly
oscillating driving action is considered and the effective Hamiltonian-like
equations for the mean motion are obtained. The respective Liouville
equation for the distribution function of the mean coordinates and momenta
is derived. \ PACS numbers 03.20.+i, 05.20.-y, 05.20.Gg
\end{abstract}

The inverted pendulum, studied by P.L. Kapitza\cite{Kapitza} and recently
revisited by S.-Y. Kim and B. Hu \cite{Kim}, is a very good example of
systems where a fast driving external action imposes a complex dynamics to
the system. The main result in Kapitza's theory is that a simple nonlinear
oscillator under the action of a rapidly oscillating force is not conserving
the mean motion on trajectories of a non perturbed analog. Instead of this,
it behaves in a completely different manner, because an additional non small
term proportional to the squared amplitude of external pulses appears in the
restoring force. This additional restoring force is responsible for the
stabilization of the inverted pendulum, which appears as a ''miracle '' and
in no way could be predicted intuitively. It is therefore very tempting to
look for such effects in extended many-body systems.

In this Letter, we propose an approach to the treatment of a large class of
such systems in a Hamiltonian-like formalism, which provides a
straightforward transition to the statistical thermodynamics of these
systems.

Consider the following dynamical equations 
\begin{eqnarray}
\stackrel{.}{\bf q}_i &=&\frac{\partial H\left( {\bf q,p,}t\right) }{%
\partial {\bf p}_i}+\frac{{\bf g}_i^{\left( 0\right) }\left( {\bf q,p}%
\right) }2+\sum_{k\geq 1}{\bf g}_i^{(k)}\left( {\bf q,p}\right) \cos \left(
k\Omega t\right) ; \\
\stackrel{.}{\bf p}_i &=&-\frac{\partial H\left( {\bf q,p,}t\right) }{%
\partial {\bf q}_i}+\frac{{\bf h}_i^{(0)}\left( {\bf q,p}\right) }2%
+\sum_{k\geq 1}{\bf h}_i^{(k)}\left( {\bf q,p}\right) \cos \left( k\Omega
t\right) ,\;\;i=1,2,\ldots ,N,
\end{eqnarray}
where ${\bf g}$ and ${\bf h}$ are arbitrary functions of all canonical
variables. Here $\Omega \gg \omega _o$, \-where $2\pi /\omega _o$ is the
characteristic time of the undriven system (with the Hamiltonian $H\left( 
{\bf q,p,}t\right) $). The series in the right-hand side of (1), (2) can be
considered as Fourier expansions of arbitrary vector functions of $t$ with
characteristic time $2\pi /\Omega $.

We do not require the system being necessarily Hamiltonian, i.e. the
equations (1), (2) are of the general type $\stackrel{.}{\bf \Gamma }={\cal F%
}({\bf \Gamma })$, where ${\bf \Gamma }\left( t\right) =\left( {\bf q,p}%
\right) $ and ${\cal F}$ is a vector function of the phase space. The
non-Hamiltonian nature of these equations means that the compressibility
does not in general vanish: 
\begin{equation}
\nabla _{{\bf \Gamma }}\cdot \stackrel{.}{\bf \Gamma }{\bf =}\nabla _{{\bf %
\Gamma }}\cdot {\cal F}\equiv \varkappa ({\bf \Gamma )\neq 0}
\end{equation}
The non-Hamiltonian dynamics is useful when considering open or driven
systems, under the action of some external influence, such as heat bath or
mechanical piston, etc. M. E. Tuckerman, C.J. Mundy and M.L. Klein \cite
{Tucker1}, \cite{Tucker2} have recently derived a correct generalization of
the Liouville equation to account for nonvanishing compressibility of phase
space (3): 
\begin{equation}
\frac{\partial \left( f\overline{J}\right) }{\partial t}=-\nabla _{{\bf %
\Gamma }}\cdot \left( \stackrel{.}{\bf \Gamma }f\overline{J}\right) ,
\end{equation}
where the Jacobian $\overline{J}$ satisfies the differential equation 
\begin{equation}
\frac{d\overline{J}}{dt}=-\overline{J\,}\nabla _{{\bf \Gamma }}\cdot 
\stackrel{.}{\bf \Gamma }.
\end{equation}
Note that substituting (5) into (4) and taking into account the obvious
relation $\frac{dJ}{dt}=\frac{\partial J}{\partial t}+\stackrel{.}{\bf %
\Gamma }\cdot \nabla _{{\bf \Gamma }}J$ leads to the conservation condition $%
df/dt=0$ or 
\begin{equation}
\frac{\partial f}{\partial t}+\stackrel{.}{\bf \Gamma }\cdot \nabla _{{\bf %
\Gamma }}f=0.
\end{equation}

The backbone of our approach is the search of solutions of eqs. (1), (2) in
the form 
\begin{eqnarray}
{\bf q}_{i} &=&{\bf Q}_{i}+\mu {\bf \chi }_{i}; \\
{\bf p}_{i} &=&{\bf P}_{i}+\mu {\bf \rho }_{i},
\end{eqnarray}
where ${\bf Q}$, ${\bf P}$ and ${\bf \chi }$, ${\bf \rho }$ are respectively
the ''slow '' and ''fast '' parts, whose characteristic times are
accordingly $T\thicksim 2\pi /\omega _{o}$ and $\tau \thicksim 2\pi /\Omega $%
, and $\mu =\omega _{o}/\Omega \ll 1$. The expression (7) was successfully
used \cite{Rabinov} for the solution of the equation of motion of inverted
pendulum. Such an ansatz (7)-(8) is physically justified, because due to
inertia the system responds weakly to fast external pulses.

Substituting (7), (8) into (1), (2) and expanding all functions in power
series of $\mu $ and retaining terms up to the first order (e.g. ${\bf g}%
_i^{(k)}({\bf q},{\bf p})\simeq {\bf g}_i^{(k)}({\bf Q},{\bf P})+\mu ({\bf %
\chi }_j\cdot \nabla _{{\bf Q}_j}){\bf g}_i^{(k)}({\bf Q},{\bf P})+\mu ({\bf %
\rho }_j\cdot \nabla _{{\bf P}_j}){\bf g}_i^{(k)}({\bf Q},{\bf P})$, etc.),
we obtain

\begin{eqnarray}
\stackrel{.}{\bf Q}_i+\mu \stackrel{.}{{\bf \chi }_i}\, &=&\frac{\partial H}{%
\partial {\bf P}_i}+\mu ({\bf \chi }_j\cdot \nabla _{{\bf Q}_j})\frac{%
\partial H}{\partial {\bf P}_i}+\mu ({\bf \rho }_j\cdot \nabla _{{\bf P}_j})%
\frac{\partial H}{\partial {\bf P}_i}+\frac{{\bf g}_i^{(0)}({\bf Q},{\bf P})}%
2+  \nonumber \\
&&\mu ({\bf \chi }_j\cdot \nabla _{{\bf Q}_j})\frac{{\bf g}_i^{(0)}({\bf Q},%
{\bf P})}2+\mu ({\bf \rho }_j\cdot \nabla _{{\bf P}_j})\frac{{\bf g}_i^{(0)}(%
{\bf Q},{\bf P})}2+\sum\limits_{k\geq 1}{\bf g}_i^{(k)}({\bf Q},{\bf P})\cos
(k\Omega t)+  \nonumber \\
&&\mu \sum\limits_{k\geq 1}({\bf \chi }_j\cdot \nabla _{{\bf Q}_j}){\bf g}%
_i^{(k)}({\bf Q},{\bf P})\cos (k\Omega t)+\mu \sum\limits_{k\geq 1}({\bf %
\rho }_j\nabla _{{\bf P}_j}){\bf g}_i^{(k)}({\bf Q},{\bf P})\cos (k\Omega t);
\end{eqnarray}
\begin{eqnarray}
\stackrel{.}{\bf P}_i+\mu \stackrel{.}{{\bf \rho }_i} &=&-\frac{\partial H}{%
\partial {\bf Q}_i}-\mu ({\bf \chi }_j\cdot \nabla _{{\bf Q}_j})\frac{%
\partial H}{\partial {\bf Q}_i}-\mu ({\bf \rho }_j\cdot \nabla _{{\bf P}_j})%
\frac{\partial H}{\partial {\bf Q}_i}+\frac{{\bf h}_i^{(0)}({\bf Q},{\bf P})}%
2+  \nonumber \\
&&\mu ({\bf \chi }_j\cdot \nabla _{{\bf Q}_j})\frac{{\bf h}_i^{(0)}({\bf Q},%
{\bf P})}2+\mu ({\bf \rho }_j\cdot \nabla _{{\bf P}_j})\frac{{\bf h}_i^{(0)}(%
{\bf Q},{\bf P})}2+\sum\limits_{k\geq 1}{\bf h}_i^{(k)}({\bf Q},{\bf P})\cos
(k\Omega t)+  \nonumber \\
&&\mu \sum\limits_{k\geq 1}({\bf \chi }_j\cdot \nabla _{{\bf Q}_j}){\bf h}%
_i^{(k)}({\bf Q},{\bf P})\cos (k\Omega t)+\mu \sum\limits_{k\geq 1}({\bf %
\rho }_j\nabla _{{\bf P}_j}){\bf h}_i^{(k)}({\bf Q},{\bf P})\cos (k\Omega t).
\end{eqnarray}
(Here we use the Einstein summation convention on repeated indices). All
functions in these equations depend on mean variables $({\bf Q},{\bf P}).$
The equations contain slow and fast terms which can be separated by
averaging over the period $\tau =2\pi /\Omega $. More precisely, (9) and
(10) are of the form: $Slow+Fast=0$. Since $\Omega \gg \omega _0$, the slow
terms may be considered constant over the time $2\pi /\Omega $ and therefore
may be substituted by their averages. The equations take the form: $\langle
Slow\rangle _\Omega +Fast\simeq 0$. On the other hand, it is evident that $%
\langle Fast\rangle _\Omega =0$ and consequently the result is a set of
coupled equations for the slow and fast variables: $\langle Slow\rangle
_\Omega =0;\,\,Fast=0$. Hence, the first step consists in integrating (9),
(10) with respect to time over the period $2\pi /\Omega $. Notice that ${\bf %
Q,P}$, as well as $H({\bf Q},{\bf P},t)$ are considered to remain constant
when integrating over ''fast time ''. Only slow terms withstand this
operation. The remaining terms form the fast equations. Particular attention
is to be given to the last and next to the last terms of (9), (10). These
terms contain the products ${\bf \chi }_j\cos (k\Omega t)$ and ${\bf \rho }%
_j\cos (k\Omega t)$. Inasmuch as ${\bf \chi }_j$ and ${\bf \rho }_j$ may
contain both $\cos (k^{\prime }\Omega t)$ and $\sin (k^{\prime }\Omega t)$
(as will become apparent later), we have here the products $\cos (k^{\prime
}\Omega t)\cos (k\Omega t)$ and $\sin (k^{\prime }\Omega t)\cos (k\Omega t)$%
. The average of the first product is nonzero for $k^{\prime }=k$, whereas
the average of the second product always vanishes. In other words, the last
two terms in (9), (10) may in general contribute both to slow and fast
equations. These latter thus read 
\begin{eqnarray}
\mu \stackrel{.}{{\bf \chi }_i}\, &=&\mu ({\bf \chi }_j\cdot \nabla _{{\bf Q}%
_j})\frac{\partial H}{\partial {\bf P}_i}+\mu ({\bf \rho }_j\cdot \nabla _{%
{\bf P}_j})\frac{\partial H}{\partial {\bf P}_i}+  \nonumber \\
&&\mu ({\bf \chi }_j\cdot \nabla _{{\bf Q}_j})\frac{{\bf g}_i^{(0)}}2+\mu (%
{\bf \rho }_j\cdot \nabla _{{\bf P}_j})\frac{{\bf g}_i^{(0)}}2%
+\sum\limits_{k\geq 1}{\bf g}_i^{(k)}\cos (k\Omega t)+  \nonumber \\
&&\mu \sum\limits_{k\geq 1}({\bf \chi }_j\cdot \nabla _{{\bf Q}_j}){\bf g}%
_i^{(k)}\cos (k\Omega t)+\mu \sum\limits_{k\geq 1}({\bf \rho }_j\nabla _{%
{\bf P}_j}){\bf g}_i^{(k)}\cos (k\Omega t);
\end{eqnarray}
\begin{eqnarray}
\mu \stackrel{.}{{\bf \rho }_i} &=&-\mu ({\bf \chi }_j\cdot \nabla _{{\bf Q}%
_j})\frac{\partial H}{\partial {\bf Q}_i}-\mu ({\bf \rho }_j\cdot \nabla _{%
{\bf P}_j})\frac{\partial H}{\partial {\bf Q}_i}+  \nonumber \\
&&\mu ({\bf \chi }_j\cdot \nabla _{{\bf Q}_j})\frac{{\bf h}_i^{(0)}}2+\mu (%
{\bf \rho }_j\cdot \nabla _{{\bf P}_j})\frac{{\bf h}_i^{(0)}}2%
+\sum\limits_{k\geq 1}{\bf h}_i^{(k)}\cos (k\Omega t)+  \nonumber \\
&&\mu \sum\limits_{k\geq 1}({\bf \chi }_j\cdot \nabla _{{\bf Q}_j}){\bf h}%
_i^{(k)}\cos (k\Omega t)+\mu \sum\limits_{k\geq 1}({\bf \rho }_j\nabla _{%
{\bf P}_j}){\bf h}_i^{(k)}\cos (k\Omega t).
\end{eqnarray}
The terms in (11), (12) do not all have the same order. While the terms $\mu 
\stackrel{.}{\bf \chi }\sim \mu \Omega {\bf \chi }\sim \omega _0{\bf \chi }$
, $\mu \stackrel{.}{\bf \rho }\sim \mu \Omega {\bf \rho }\sim \omega _0{\bf %
\rho }$ are not small, the terms with $\mu {\bf \chi }$, $\mu {\bf \rho }$
are much smaller. So, we find for the fast equations accurate up to
zero-order terms in $\mu $ 
\begin{equation}
\mu \stackrel{\cdot }{{\bf \chi }_i}\simeq \sum\limits_{k\geq 1}{\bf g}%
_i^{(k)}\cos (k\Omega t);\;\;\;\mu \stackrel{\cdot }{{\bf \rho }_i\,}\simeq
\sum\limits_{k\geq 1}{\bf h}_i^{(k)}\cos (k\Omega t).
\end{equation}
Integrating these equations, we obtain 
\begin{equation}
{\bf \chi }_i\simeq \frac 1{\mu \Omega }\sum\limits_{k\geq 1}\frac{{\bf g}%
_i^{(k)}}k\sin (k\Omega t);\;\;\;{\bf \rho }_i\simeq \frac 1{\mu \Omega }%
\sum\limits_{k\geq 1}\frac{{\bf h}_i^{(k)}}k\sin \left( k\Omega t\right) .
\end{equation}
Subsequently, we solve the fast equations taking into account the terms of
first-order in $\mu $ too, substituting (14) into (11), (12) and again
integrating with respect to time (recall that ${\bf Q}_i,\,{\bf P}_i,\,H(%
{\bf Q},{\bf P},t)$ are held constant) 
\begin{eqnarray}
{\bf \chi }_i &\simeq &\frac 1{\mu \Omega }\sum\limits_{k\geq 1}\frac{{\bf g}%
_i^{(k)}}k\sin (k\Omega t)-\frac 1{\mu \Omega ^2}\sum\limits_{k\geq 1}\frac 1%
{k^2}\left( {\bf g}_j^{(k)}\cdot \nabla _{{\bf Q}_j}+{\bf h}_j^{(k)}\cdot
\nabla _{{\bf P}_j}\right) \cdot (\frac{\partial H}{\partial {\bf P}_i}+%
\frac{{\bf g}_i^{(o)}}2)\cdot \cos (k\Omega t)-  \nonumber \\
&&\frac 1{\mu \Omega ^2}\,\sum\limits_{l,k\geq 1(l\neq k)}\frac 1{2k(k-l)}%
\left( {\bf g}_j^{(k)}\cdot \nabla _{{\bf Q}_j}+{\bf h}_j^{(k)}\cdot \nabla
_{{\bf P}_j}\right) {\bf g}_i^{(l)}\cos [(k-l)\Omega t]-  \nonumber \\
&&\frac 1{\mu \Omega ^2}\sum\limits_{l,k\geq 1}\frac 1{2k(k+l)}\left( {\bf g}%
_j^{(k)}\cdot \nabla _{{\bf Q}_j}+{\bf h}_j^{(k)}\cdot \nabla _{{\bf P}%
_j}\right) {\bf g}_i^{(l)}\cos [(k+l)\Omega t];
\end{eqnarray}
\begin{eqnarray}
{\bf \rho }_i &\simeq &\frac 1{\mu \Omega }\sum\limits_{k\geq 1}\frac{{\bf h}%
_i^{(k)}}k\sin (k\Omega t)+\frac 1{\mu \Omega ^2}\sum\limits_{k\geq 1}\frac 1%
{k^2}\left( {\bf g}_j^{(k)}\cdot \nabla _{{\bf Q}_j}+{\bf h}_j^{(k)}\cdot
\nabla _{{\bf P}_j}\right) \cdot (\frac{\partial H}{\partial {\bf Q}_i}-%
\frac{{\bf h}_i^{(0)}}2)\cdot \cos (k\Omega t)-  \nonumber \\
&&\frac 1{\mu \Omega ^2}\,\sum\limits_{l,k\geq 1(l\neq k)}\frac 1{2k(k-l)}%
\left( {\bf g}_j^{(k)}\cdot \nabla _{{\bf Q}_j}+{\bf h}_j^{(k)}\cdot \nabla
_{{\bf P}_j}\right) {\bf h}_i^{(l)}\cos [(k-l)\Omega t]-  \nonumber \\
&&\frac 1{\mu \Omega ^2}\sum\limits_{l,k\geq 1}\frac 1{2k(k+l)}\left( {\bf g}%
_j^{(k)}\cdot \nabla _{{\bf Q}_j}+{\bf h}_j^{(k)}\cdot \nabla _{{\bf P}%
_j}\right) {\bf h}_i^{(l)}\cos [(k+l)\Omega t].
\end{eqnarray}

Expressions (15), (16) represent a rapidly oscillating response of
coordinates and momenta of the system to the fast external actions. The
point is that it is not an unique reaction of the system. The mean
trajectories $({\bf Q},{\bf P})$ are altered too. Inserting (15), (16) into
(9), (10) and averaging over the period $2\pi /\Omega $, we obtain the
following equations of motion for the mean canonical coordinates: 
\begin{eqnarray}
\stackrel{.}{\bf Q}_{i} &=&\,\left( \frac{\partial H}{\partial {\bf P}_{i}}+%
\frac{{\bf g}_{i}^{(0)}}{2}\right) -  \nonumber \\
&&\frac{1}{2\Omega ^{2}}\sum_{k\geq 1}\frac{1}{k^{2}}\left\{ \left( {\bf g}%
_{j^{\prime }}^{(k)}\cdot \nabla _{{\bf Q}_{j\acute{}}}+{\bf h}_{j\acute{}%
}^{(k)}\cdot \nabla _{{\bf P}_{j}\acute{}}\right) \left[ \left( \frac{%
\partial H}{\partial {\bf P}_{j}}+\frac{{\bf g}_{j}^{(0)}}{2}\right) \cdot
\nabla _{{\bf Q}_{j}}-\left( \frac{\partial H}{\partial {\bf Q}_{j}}-\frac{%
{\bf h}_{j}^{(0)}}{2}\right) \cdot \nabla _{{\bf P}_{j}}\right] \right\} \;%
{\bf g}_{i}^{(k)}-  \nonumber \\
&&\frac{1}{2\Omega ^{2}}\sum_{l\neq k\geq 1}\frac{1}{2k\left( l-k\right) }%
\left\{ \left( {\bf g}_{j\acute{}}^{(l)}\cdot \nabla _{{\bf Q}_{j\acute{}}}+%
{\bf h}_{j\acute{}}^{(l)}\cdot \nabla _{{\bf P}_{j}\acute{}}\right) \left( 
{\bf g}_{j}^{(k)}\cdot \nabla _{{\bf Q}_{j}}+{\bf h}_{j}^{(k)}\cdot \nabla _{%
{\bf P}_{j}}\right) \right\} \;{\bf g}_{i}^{(\left| l-k\right| )}-  \nonumber
\\
&&\frac{1}{2\Omega ^{2}}\sum_{l,k\geq 1}\frac{1}{2k\left( l+k\right) }%
\left\{ \left( {\bf g}_{j\acute{}}^{(l)}\cdot \nabla _{{\bf Q}_{j\acute{}}}+%
{\bf h}_{j\acute{}}^{(l)}\cdot \nabla _{{\bf P}_{j}\acute{}}\right) \left( 
{\bf g}_{j}^{(k)}\cdot \nabla _{{\bf Q}_{j}}+{\bf h}_{j}^{(k)}\cdot \nabla _{%
{\bf P}_{j}}\right) \right\} {\bf g}_{i}^{(l+k)};
\end{eqnarray}
\begin{eqnarray}
\stackrel{.}{\bf P}_{i} &=&-\left( \frac{\partial H}{\partial {\bf Q}_{i}}-%
\frac{{\bf h}_{i}^{(0)}}{2}\right) -  \nonumber \\
&&\frac{1}{2\Omega ^{2}}\sum_{k\geq 1}\frac{1}{k^{2}}\left\{ \left( {\bf g}%
_{j\acute{}}^{(k)}\cdot \nabla _{{\bf Q}_{j\acute{}}}+{\bf h}_{j\acute{}%
}^{(k)}\cdot \nabla _{{\bf P}_{j}\acute{}}\right) \left[ \left( \frac{%
\partial H}{\partial {\bf P}_{j}}+\frac{{\bf g}_{j}^{(0)}}{2}\right) \cdot
\nabla _{{\bf Q}_{j}}-\left( \frac{\partial H}{\partial {\bf Q}_{j}}-\frac{%
{\bf h}_{j}^{(0)}}{2}\right) \cdot \nabla _{{\bf P}_{j}}\right] \right\} 
{\bf h}_{i}^{(k)}-  \nonumber \\
&&\frac{1}{2\Omega ^{2}}\sum_{l\neq k\geq 1}\frac{1}{2k\left( l-k\right) }%
\left\{ \left( {\bf g}_{j\acute{}}^{(l)}\cdot \nabla _{{\bf Q}_{j\acute{}}}+%
{\bf h}_{j\acute{}}^{(l)}\cdot \nabla _{{\bf P}_{j}\acute{}}\right) \left( 
{\bf g}_{j}^{(k)}\cdot \nabla _{{\bf Q}_{j}}+{\bf h}_{j}^{(k)}\cdot \nabla _{%
{\bf P}_{j}}\right) \right\} \;{\bf h}_{i}^{(\left| l-k\right| )}-  \nonumber
\\
&&\frac{1}{2\Omega ^{2}}\sum_{l,k\geq 1}\frac{1}{2k\left( l+k\right) }%
\left\{ \left( {\bf g}_{j\acute{}}^{(l)}\cdot \nabla _{{\bf Q}_{j\acute{}}}+%
{\bf h}_{j\acute{}}^{(l)}\cdot \nabla _{{\bf P}_{j}\acute{}}\right) \left( 
{\bf g}_{j}^{(k)}\cdot \nabla _{{\bf Q}_{j}}+{\bf h}_{j}^{(k)}\cdot \nabla _{%
{\bf P}_{j}}\right) \right\} {\bf h}_{i}^{(l+k)}.
\end{eqnarray}
It is pertinent to note that all additional terms ($\sim 1/\Omega ^{2}$) in
these equations come from the last two terms of (9), (10) when integrating
of the function $\cos (k^{\prime }\Omega t)\cos (k\Omega t)$ over time: $%
\frac{\Omega }{2\pi }\int\limits_{0}^{2\pi /\Omega }\cos (k^{\prime }\Omega
t)\cos (k\Omega t)dt=\frac{1}{2}\delta _{kk^{\prime }}$.

Relations (17), (18) constitute the effective equations of motion of a
system subject to high-frequency external actions. The respective Liouville
equation for the effective distribution function $F({\bf Q},{\bf P,}t)$
follows immediately from (17), (18) in the form (4)-(5): 
\begin{equation}
\frac{\partial \left( F\overline{{\cal J}}\right) }{\partial t}=-\nabla _{%
{\cal G}}\cdot \left( \stackrel{.}{\cal G}F\overline{{\cal J}}\right) ;\;\;\;%
\frac{d\overline{{\cal J}}}{dt}=-\overline{{\cal J}}\,\nabla _{{\cal G}%
}\cdot \stackrel{.}{\cal G};
\end{equation}
where ${\cal G}=\left( {\bf Q},{\bf P}\right) $ is the phase space vector of
the mean coordinates, $\overline{{\cal J}}$ is the inverse Jacobian of the
transformation from the initial mean phase coordinates to the coordinates at
time $t$: 
\begin{equation}
\overline{{\cal J}}=\frac{\partial \left( {\bf Q}_1^{(0)},...,{\bf Q}%
_N^{(0)},{\bf P}_1^{(0)},...,{\bf P}_N^{(0)}\right) }{\partial \left( {\bf Q}%
_1^{(t)},...,{\bf Q}_N^{(t)},{\bf P}_1^{(t)},...,{\bf P}_N^{(t)}\right) }.
\end{equation}
The phase space velocity vector$\stackrel{.}{\text{ }{\cal G}}$ is given by
the right-hand sides of the equations (17), (18).

In the simple case of a forced nonlinear oscillator with a Hamiltonian $%
H=p^2/2m+U(q)$ and the driving force $F(q)\cos \Omega t$, we have the only
nonvanishing amplitude $h^{(1)}(q)=F(q)$ and from the equations (17), (18)
follows the equation of motion in the mean coordinate 
\begin{equation}
m\stackrel{..}{Q}+\,\left( \frac{dU}{dq}\right) _Q+\frac 12\frac{F(Q)}{%
\Omega ^2}\left( \frac{dF}{dq}\right) _Q=0.
\end{equation}
This coincides with the result first obtained by Kapitza \cite{Kapitza}, 
\cite{Rabinov} . The last term in (21) stabilizes the inverted pendulum. In
other cases, there may be different extra terms in equations (17), (18) and
respectively in the Liouville equation. These terms depend on various
combinations of derivatives of the functions ${\bf g}_i^{(k)},{\bf h}%
_i^{(k)} $ with respect to canonical variables and may be responsible for
many interesting dynamical and thermodynamic effects.

It is important to emphasize that even in the case when the equations (1),
(2) are Hamiltonian, the equations (17), (18) may be non-Hamiltonian, i.e.
the compressibility does not in general vanish $\varkappa \left( {\cal G}%
\right) {\bf =\nabla _{{\cal G}}\stackrel{.}{\cal G}\neq 0}$. It is not
surprising since a driven system behaves like a thermodynamically open
system.

We would like to thank Dr. M.E. Tuckerman for helpful correspondence on the
subject matter of this paper. This work was supported by Conselho Nacional
de Desenvolvimento Cient\'{i}fico e Tecnol\'{o}gico (CNPq, Brazil).

\end{document}